\documentclass[osajnl,twocolumn,showpacs,superscriptaddress,10pt]{revtex4-1} 
\usepackage{amsmath,amssymb,graphicx}
\begin{document}

\title{Frozen photons  in Jaynes Cummings arrays}

\author{Nikolaos Schetakis}\email{Corresponding author: nsxetakis@yahoo.gr}
\affiliation{Science Department, Technical University of Crete, Chania, Crete, Greece, 73100}

\author{Thomas Grujic}\email{Corresponding author: t.grujic1@physics.ox.ac.uk}
\affiliation{Clarendon Laboratory, University of Oxford, Parks Road, Oxford OX1 3PU, UK}

\author{Stephen Clark}
\affiliation{Clarendon Laboratory, University of Oxford, Parks Road, Oxford OX1 3PU, UK}
\affiliation{Centre for Quantum Technologies, National University of Singapore, 2 Science Drive 3, Singapore 117542}

\author{Dieter Jaksch}
\affiliation{Clarendon Laboratory, University of Oxford, Parks Road, Oxford OX1 3PU, UK}
\affiliation{Centre for Quantum Technologies, National University of Singapore, 2 Science Drive 3, Singapore 117542}

\author{Dimitris Angelakis}\email{Corresponding author: dimitris.angelakis@gmail.com}
\affiliation{Science Department, Technical University of Crete, Chania, Crete, Greece, 73100}
\affiliation{Centre for Quantum Technologies, National University of Singapore, 2 Science Drive 3, Singapore 117542}

\begin{abstract}
We  study the origin  of `frozen' states in coupled  Jaynes-Cummings-Hubbard arrays  in the presence of losses. For the case of half the array initially populated with photons while the other half is left empty we show the emergence of self-localized photon  or `frozen' states for specific  values of the local atom-photon coupling.  We analyze the dynamics in the quantum regime  and discover important additional features appear not captured by a semiclassical treatment, which we  analyze  for different array sizes and filling fractions. We trace the origin of this interaction-induced photon `freezing' to the suppression of excitation of propagating modes in the system at large interaction strengths. We discuss in detail the possibility to experimentally probe the relevant transition by analyzing the emitted photon correlations. We find a strong signature of the effect in the emitted photons statistics.
\end{abstract}


\maketitle 

\section{Introduction}
The years since the first proposals to realise strongly correlated many-body states of light in resonator array architectures \cite{angelakis2007photon, hartmann2006strongly, greentree2006quantum} have seen a rapid growth of interest in these structures along several lines of investigation. Perhaps most recently, there has been significant attention paid to finding novel observable non-equilibrium dynamical effects in modest-sized arrays, both in the steady state of driven-dissipative systems \cite{tomadin2010signatures, bamba2011origin, leib2010bose, carusotto2009fermionized, gerace2009quantum, grujic2012repulsively}, and in explorations of coherent array dynamics \cite{makin2009time, schmidt2010nonequilibrium,longo2011few,wong2011two,hartmann2008migration}. Resonator arrays are in many ways ideal platforms for the exploration of non-equilibrium quantum phenomena, allowing relative ease of access to dynamical observables via localised measurements of photon fields. 

In this work we investigate the time evolution of nonlinear arrays, going beyond the one- or two- photon limit into a strongly-correlated many-body regime. We find evidence for an interaction-induced `freezing' of domain walls of photons in initially half-filled one dimensional array systems. The resonator nonlinearity must be sufficiently large for localisation effects to set in, beyond which the photon population remains trapped in half of the system. We show that a semiclassical treatment similar to that first presented in \cite{schmidt2010nonequilibrium} for the limiting case of two coupled resonators predicts a sharp transition between localised domain formation and delocalised dynamics in which photons tunnel between both halves of the system. 
\begin{figure}[htbp]
\centerline{\includegraphics[width=.95\columnwidth]{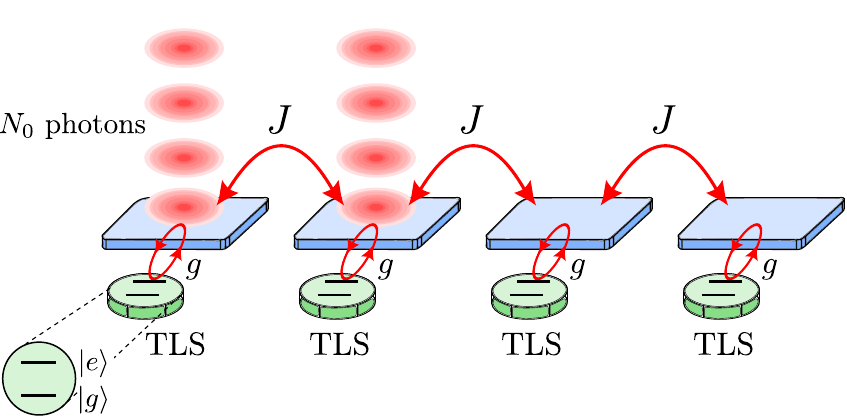}}
\caption{Schematic of our system and the initial conditions considered. The left half of a one dimensional array with $M$ resonators (here $M=4$) is initialised in a Fock state of $N_0$ photons (in this particular schematic, $N_0=4$). Each resonator is coherently coupled to its two nearest neighbours with associated tunnelling rate $J$. Each resonator is also coherently coupled to a two-level system (TLS) with Jaynes-Cummings coupling parameter $g$. }
\label{fig:sys_schematic}
\end{figure}

Going beyond the semiclassical approach, fully quantum calculations confirm the frozen photon dynamics for strongly nonlinear arrays while also revealing features not present in the semiclassical calculation. 

Our findings can be seen in some ways as a photonic analog of the non-equilibrium dynamics of XXZ chains where it has been recently shown that strong nearest-neighbor interactions can lead to the formation of polarised domains which strongly influence the transport properties of spin chains \cite{mendoza2013dephasing}. These ferromagnetic domains have been shown to be stable. They are spectrally separated from mobile states of the system which are capable of breaking them. 

We show a related but distinct photonic equivalent. Our `domains' of photons remain trapped over timescales large with respect to characteristic system rates due to a vanishing overlap between the initial pumped states we consider, and propagating modes of the system in the limit of large local interactions. 

\section{System}
The system we consider is a one-dimensional linear array of $M$ coupled optical resonators. Each resonator features a relevant mode of frequency $\omega_r$, and is coherently coupled to its nearest neighbours, as shown schematically in Fig.~\ref{fig:sys_schematic}. A single two-level system (TLS) with transition frequency $\omega_a$ is coherently coupled via a Jaynes-Cummings interaction to each resonator, with coupling strength $g$. In this work, we consider only the on-resonance case $\Delta = \omega_r - \omega_a = 0$. The governing Hamiltonian is then the well-known Jaynes-Cummings-Hubbard (JCH) Hamiltonian:
\begin{eqnarray}
\hat{\mathcal{H}} & = & \sum_{j}\left[(  \omega_r \hat{a}^\dag_j \hat{a}_j + \omega_{a} \hat{\sigma}^+_j \hat{\sigma}_j^- + g \left ( \hat{a}^\dag_j \hat{\sigma}^-_j + \hat{a}_j \hat{\sigma}^+_j\right )\right ]\nonumber\\
& - & J \sum_{\langle j, j'\rangle} \hat{a}^\dag_j \hat{a}_{j'}. 
\label{eq:jch_ham}
\end{eqnarray}
Here $\hat{a}_j$ is the photon destruction operator for resonator $j$, and the $\hat{\sigma}^\pm_j$ are the raising / lowering operators for the  TLS coupled to resonator $j$. The set of nearest neighbour resonators is denoted by $\langle j,j' \rangle$. The Hamiltonian $\hat{\mathcal{H}}$ commutes with the total excitation number operator 
\begin{equation}
\hat{\mathcal{N}} = \sum_j \left ( \hat{a}_j^\dag \hat{a}_j + \hat{\sigma}^+_j \hat{\sigma}^-_j \right ). 
\end{equation}

Throughout this work we shall consider a specific type of initial state. Namely, we initialise one half of the system in a Fock state of $N_0$ photons, with each TLS in its ground state. 
\begin{equation}
| \Psi (0) \rangle = \prod _{j=1}^{M/2} |g, N_0\rangle_j \otimes \prod_{j=M/2+1}^M |g,0 \rangle_j,
\label{eq:init_state}
\end{equation}
where $|g, n \rangle_j$ denotes a photonic Fock state of $n$ photons in resonator $j$.

Generalising \cite{schmidt2010nonequilibrium} to the case of an extended system, of central interest in the following analysis will be the photon imbalance $Z(t)$ between the left (L) and right (R) halves of the system, as defined by:
\begin{equation}
Z(t) = \frac{\sum_{j=1}^{M/2} \langle \hat{a}_j^\dag \hat{a}_j \rangle(t) - \sum_{j=M/2+1}^{M} \langle \hat{a}_j^\dag \hat{a}_j \rangle(t)}{\sum_{j=1}^{M} \langle \hat{a}_j^\dag \hat{a}_j \rangle(t)}
\end{equation}
In particular, we find that time-averaging $Z(t)$ neatly encapsulates details of the photon dynamics. We denote such time averages in the following by $\bar{Z}$. Values of $\bar{Z}$ close to zero imply delocalisation of photons across the two halves of the system, either oscillating back and forth in some manner or reaching an approximately even distribution. Meanwhile $\bar{Z} \approx 1$ implies a photon population trapped in one side of the system for a substantial period of time. 

\section{Semi-classical treatment}
We begin our analysis of the dynamics of the system Eq.~(\ref{eq:jch_ham}) at the semiclassical level, thereby making contact with the related previous work of \cite{schmidt2010nonequilibrium}. We first use the Heisenberg equation of motion $\frac{d}{dt}\hat{O} = i [\hat{\mathcal{H}}, \hat{O}]$ to generate evolution equations for the photonic and TLS operator expectation values. As the semi-classical approximation entails factorising the expectation values of operator products into products of expectation values (e.g. $\langle \hat{a}^\dag_j \hat{\sigma}^-_j \rangle = \langle \hat{a}^\dag_j \rangle \langle \hat{\sigma}^-_j \rangle$), we need only generate equations for the three operators $\hat{a}_j, \hat{\sigma}^-_j$, $\hat{\sigma}^z_j$. 

Defining $(\alpha_j, m_j, z_j) \equiv (\langle \hat{a}_j \rangle, \langle \hat{\sigma}^-_j \rangle, \langle \hat{\sigma}^z_j \rangle)$, we obtain the set of coupled differential equations for their evolution:
\begin{eqnarray}
\dot{\alpha_j} & = & -i \omega_r \alpha_j - i g m_j + i J ((1- \delta_{j, 1}) \alpha_{j-1} + (1- \delta_{j, M}) \alpha_{j+1})\nonumber\\
\dot{m_j} & = & -2i \omega_a m_j + i g \alpha_j z_j \nonumber\\
\dot{z_j} & = & -2i g \left ( \alpha_j m_j^* - \alpha_j^* m_j \right ),
\label{eq:SC}
\end{eqnarray}
where the delta functions take into account the open boundary conditions. 

The initial conditions corresponding to our chosen state of Eq.~(\ref{eq:init_state}) are 
\begin{eqnarray}
\alpha_j & = & \left\{
     \begin{array}{lr}
       \sqrt{N_0} : j \le \frac{M}{2}\\
       0 : j > M/2
     \end{array}
   \right.,\nonumber\\
z_j & = & -1,\nonumber\\
m_j & = & 0. 
\end{eqnarray}
We note that it has been shown that for the case of $M=2$ resonators, a qualitative change in the population imbalance dynamics occurs sharply at a critical coupling $g_c \approx 2.8 \sqrt{N_0} J$ \cite{schmidt2010nonequilibrium}. Specifically, for $g < g_c$, photons move between the resonators with a characteristic tunnelling time. Around $g = g_c$, this period diverges, leading to a `self-trapped' regime for $g > g_c$ in which the photon population remains localised in one resonator. 

\begin{figure}[htbp]
\centerline{\includegraphics[width=.8\columnwidth]{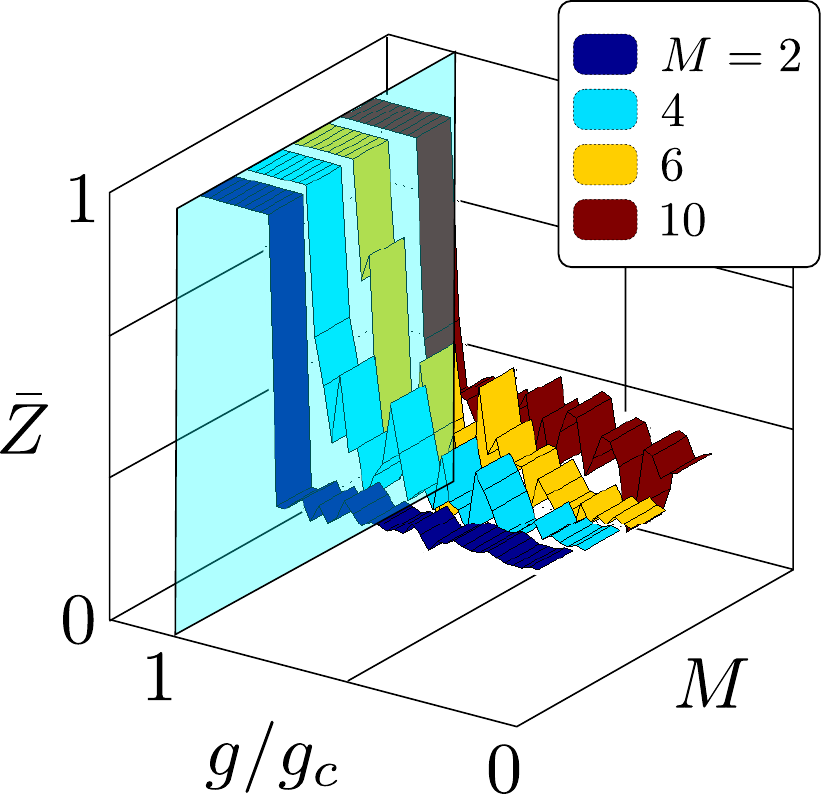}}
\caption{The time-averaged photon imbalance $\bar{Z}$ according to a semi-classical treatment of an $M$-resonator array. $\bar{Z}$ denotes the average of $Z(t)$ over a time interval $t J \in [0, 20]$. The transparent blue plane marks the critical coupling predicted by semi-classical theory for the case of $M=2$ resonators \cite{schmidt2010nonequilibrium}. }
\label{fig:semiclassical}
\end{figure}

With these results in mind, we look now at the time averaged photon imbalance $\bar{Z}$ for larger arrays as calculated by time evolving the set of Eqns.~(\ref{eq:SC}), shown in Fig.~\ref{fig:semiclassical}. We see that semi-classical theory still predicts a sharp localised / delocalized transition at the critical coupling $g = g_c$, regardless of the system size $M$. 

\section{The fully quantum regime}
Going now beyond semiclassical (SC) theory, which can only be valid in the limit of large number of excitations, we investigate whether an analogue of the localisation predicted by the SC equations persists in the fully quantum regime of a few ($N_0 \le 4$) excitations. Explicitly constructing a matrix representation of the Hamiltonian of Eq.~(\ref{eq:jch_ham}) and time evolving by applying the unitary operator $\mathcal{U}(t) = \exp(-i \mathcal{H} t)$ to the initial state $|\Psi(t) \rangle$ becomes numerically challenging beyond even $M=2$ resonators. The two-species nature of the Jaynes-Cummings Hubbard Hamiltonian, coupled with the necessity of retaining a sufficient number of photons per resonator in calculations so as to avoid truncation error lead to a large Hilbert space dimension. Some progress is possible by projecting the dynamics into fixed particle number subspaces, however we turn instead to a compact matrix product state (MPS) representation of the wave function \cite{perez2007matrix}. This representation is ideally suited for representing the state of one-dimensional systems with at most nearest-neighbor couplings as in our case. Efficient and accurate Hamiltonian evolution of the MPS is achieved via the time-evolving block decimation (TEBD) algorithm \cite{vidal2003efficient}. 

\begin{figure}[htbp]
\centerline{\includegraphics[width=.96\columnwidth]{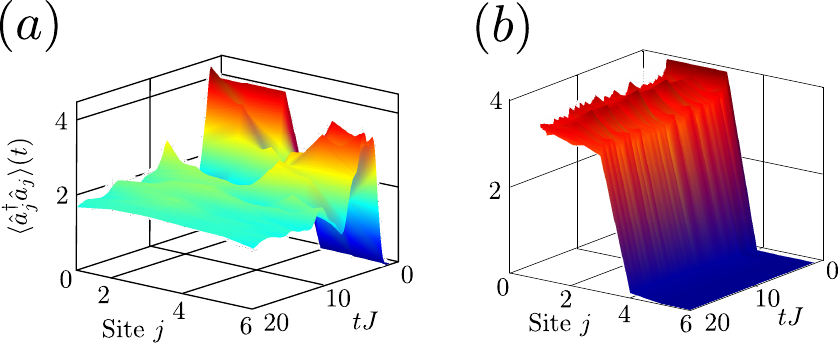}}
\caption{Photon number evolution for an $M=6$ resonator system with the first three resonators pumped with $N_0=4$ photon Fock states at $t=0$. (a) A weak nonlinearity $g = 0.1 J$.  (b) Strong Jaynes-Cummings nonlinearity $g = 15J$. The apparent reduction in the mean photon number per site over time is accounted for by the TLS excitation -- we have checked that the total excitation in the system is preserved by our numerics to $1\%$ over the simulation interval. Simulation parameters: all calculations kept a minimum of $n_{\rm max} = {\rm min} (N_0 M / 2, 7)$ photons per resonator in the computational basis. A matrix product state truncation parameter of $\chi = 100$ was found to be sufficient to avoid cumulative errors. }
\label{fig:M_6_phot_num_evolution}
\end{figure}

Figure~\ref{fig:M_6_phot_num_evolution} shows TEBD simulations of the local photon density in an $M=6$ resonator array with the left half initially pumped with Fock states of $N_0=4$ photons, for two Jaynes-Cummings nonlinearities, one weak and one strong (relative to the photon tunnelling rate $J$). For weak nonlinearities photons initially oscillate between the two halves of the system, reflecting from the boundaries and eventually leading to a uniformly distributed population, i.e. zero photon imbalance $\bar{Z} \approx 0$. 

For arrays with strong nonlinearities on the other hand, such as shown in Figure~\ref{fig:M_6_phot_num_evolution}~(b), photons remain essentially trapped in the left-hand side of the system for very long times ($>98\%$ of the photon population remains in the first three sites over the simulation window). The self-trapping phenomenon predicted by semiclassical theory seems then to persist in the fully quantum regime of few photons. 

\begin{figure}[htbp]
\centerline{\includegraphics[width=.96\columnwidth]{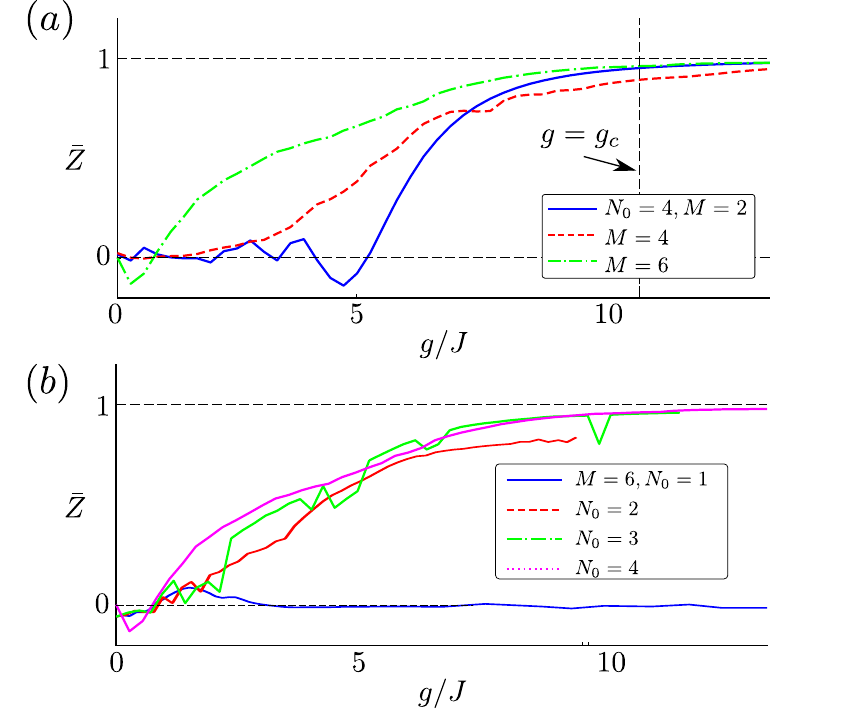}}
\caption{An exploration of the time averaged photon imbalance $\bar{Z}$. (a) As a function of the system size $M$, holding the initial excitation $N_0$ fixed. (b) $\bar{Z}$ as function of the initial excitation $N_0$, holding instead the system size fixed. }
\label{fig:time_averaged_photon_imbalance_varying_N0_and_M}
\end{figure}

Figure~\ref{fig:time_averaged_photon_imbalance_varying_N0_and_M} however characterises the emergence of these domains of `frozen' photons, showing that semiclassical theory is insufficient to fully capture all qualitative details of the effect in the low-excitation regime. Figure~\ref{fig:time_averaged_photon_imbalance_varying_N0_and_M}~(a) shows the results of rigorous TEBD simulations for arrays of different sizes, pumped with different numbers of initial photons. We see that the `transition' between delocalized ($\bar{Z} \approx 0$) and localised ($\bar{Z} \approx 1$) dynamics becomes broader with increasing system size $M$, and never resembles the sharp semi-classical transition of Fig.~\ref{fig:semiclassical}. The qualitative trend, however, towards localisation with increasing nonlinearity $g$ occurs irrespective of $M$. Figure~\ref{fig:time_averaged_photon_imbalance_varying_N0_and_M}~(b) meanwhile shows that for $N_0>1$, the magnitude of the initial excitation does not significantly affect the rate at which the system approaches the `frozen' regime. Interestingly however we find that the case of a single photon pumped into the left half of the system never exhibits localisation behaviour, no matter how large the Jaynes-Cummings nonlinearity $g$. 

\begin{figure}[htbp]
\centerline{\includegraphics[width=.98\columnwidth]{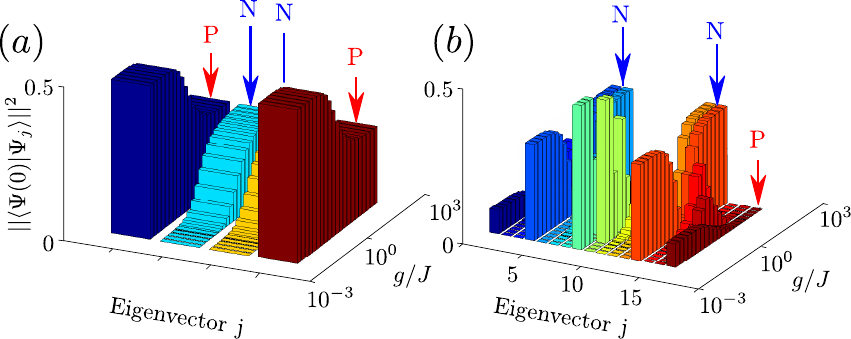}}
\caption{The projection of the initial state $|\Psi (t=0)\rangle$ into each of the eigenstates spanning the subspace consistent with the total number of excitation in the system, $N_0M/2$, for the simplest case of a dimer of $M=2$ resonators. (a) Shows overlaps for an initial excitation of just $N_0=1$ photon in the first resonator. (b) The corresponding projections for an initial state of $N_0=4$ photons. Both plots show the evolution of the different projections as the Jaynes-Cummings nonlinearity $g$ is ramped up. Eigenmodes marked `P' (for `propagating') have a nonzero photon correlation function C across the two halves of the system. Modes marked `N' have vanishing $C$ as $g$ increases. Note that the larger dimension of the subspace for $N_0 = 4$ results in more eigenstates in (b). }
\label{fig:projections_initial_states}
\end{figure}

We can understand this qualitative difference in behaviour between $N_0 = 1$ and $N_0 > 1$ by examining the initial state $|\Psi(0) \rangle$ in the eigenbasis of the Hamiltonian $\hat{\mathcal{H}}$. We work in the Hilbert subspace spanned by eigenvectors $\{ \Psi_j \}$ commuting with the total excitation operator $\hat{\mathcal{N}}$ with eigenvalue $n = N_0 M / 2$, and calculate the overlap of the initial state with each of these eigenstates $||\langle\Psi(0) | \Psi_j \rangle||^2$. These overlaps are presented in Fig.~\ref{fig:projections_initial_states} for a minimal `array' of $M=2$ resonators with both $N_0 = 1$ and $N_0 = 4$ initial photons, as a function of the nonlinearity $g$. As expected, we see a shuffling of excitation between the various eigenmodes as $g$ changes. The properties of the eigenmodes having significant overlap with the initial state for a given nonlinearity $g$ determine the properties of the time evolution of the system. Of particular relevance is a measure of the `photon current' through the centre of the system, or alternatively the degree of photon delocalization across the halves of the array. Both these quantities are reflected in a finite value of the expectation value $C = | \langle \hat{a}^\dag_{M/2} \hat{a}_{M/2+1} \rangle |$. On measuring $C$ for each of the modes $|\Psi_j \rangle$, we find that the initial state for the case of a single pumped initial photon $N_0 = 1$ has a finite overlap with `current-carrying' modes even in the limit of a large nonlinearity $g$. In contrast, only non-propagating modes are substantially excited for larger excitations $N_0 > 1$, leading to the frozen domains of Fig.~\ref{fig:M_6_phot_num_evolution}~(b). 

\section{Probing the frozen dynamics in an experiment}
\begin{figure}[htbp]
\centerline{\includegraphics[width=.98\columnwidth]{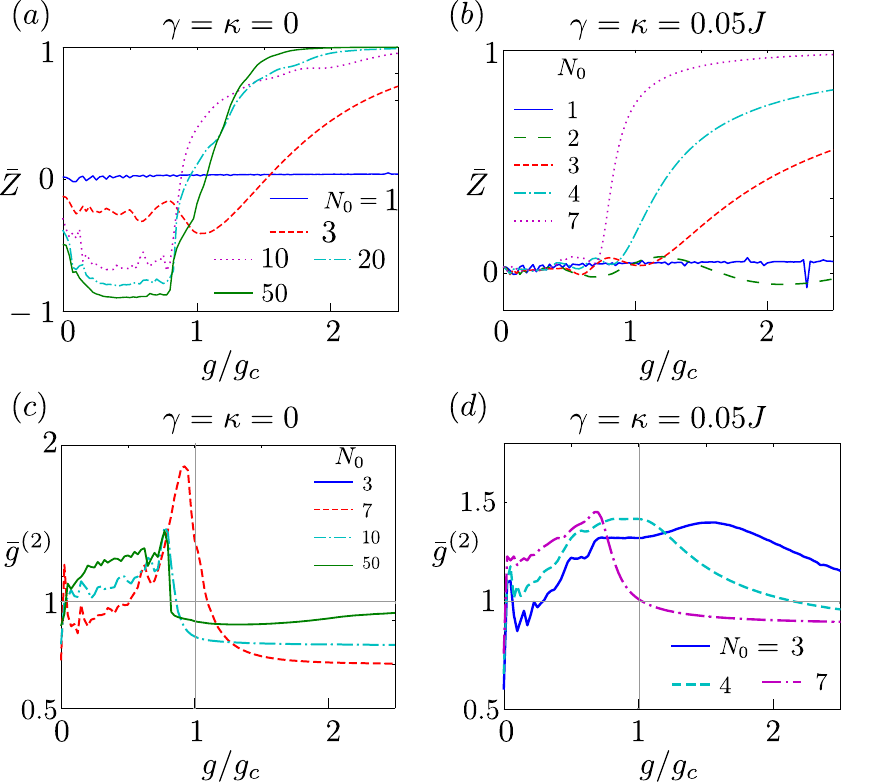}}
\caption{Charting the localisation - delocalization transition for a minimal $M=2$ resonator array through both the time averaged photon imbalance $\bar{Z}$ (top row) and the time-averaged photon correlator $\bar{g}^{(2)}$ (bottom row) for increasing initial photon number $N_0$. $\bar{Z}$ and $\bar{g}^{(2)}$ are calculated for systems with no losses (left column), and for small finite loss rates $\gamma = \kappa = 0.05 J$. }
\label{fig:with_loss_stuff}
\end{figure}

Finally we present calculations showing experimentally relevant photonic observables which give signatures of the transition between localised and delocalized dynamics. While the photon number imbalance $\bar{Z}$ between the two halves of the system may be measurable via quantum non-demolition measurements on frequency shifts of the TLSs, we focus here on purely photonic observables that can be extracted from the emitted photons from the structure. In particular, we find that measurement of the local second order photon correlations $g^{(2)}_L = \langle \hat{a}^\dag_j \hat{a}^\dag_j \hat{a}_j \hat{a}_j \rangle / \langle \hat{a}^\dag_j \hat{a}_j\rangle^2$ yields signatures of the transition. Figure.~\ref{fig:with_loss_stuff} shows that the freezing of population in one half of the system (where the photon imbalance is $\bar{Z} \approx 1$) is accompanied by a qualitative change in the on-site correlations from $g^{(2)} > 1$ to $g^{(2)} \approx 1$. Figures.~\ref{fig:with_loss_stuff} (a) and (c) show that both the photon imbalance and the time-averaged correlator $g^{(2)}$ approach a limiting behaviour as the initial number of photons grows large, with a sharp transition in observables in the vicinity of the critical point predicted by semiclassical theory $g = g_c$. 

Experimentally, the photon statistics encoded in $g^{(2)}$ are mapped on to the statistics of photons leaking from resonators with finite line widths as characterised by a loss rate $\gamma$. To assess whether the correlator $g^{(2)}$ can serve as a probe of the transition in realistic settings with a finite resonator loss rate, we include Markovian photon loss processes at rate $\gamma$ and TLS de-excitation at rate $\kappa$ via a quantum master equation formalism, time evolving the system density matrix $\rho$ from the initial state of Eq.~(\ref{eq:init_state}) under the evolution:
\begin{equation}
\dot\rho(t)=-i[H,\rho]  +\sum_{i=L,R} (\kappa {\cal L}   [ a_i  ] +\gamma {\cal L}  [\sigma_i^{-}]),
\label{eq:master}
\end{equation}
where the action of the dissipator $\mathcal{L}$ is defined as $\cal L$=$( 2 O\rho O^\dagger - O^\dagger O\rho - \rho O^\dagger O)/2$. Figures.~\ref{fig:with_loss_stuff} (b) and (d) show that the introduction of a finite loss rate acts to smear the transition, pushing localisation to larger nonlinearities $g$. However for sufficiently large initial photon pumping (around $N_0\approx 7$), the statistics of the emitted photons can be used to infer a change from delocalised physics (characterised by $\bar{g}^{(2)} > 1$) to the localised case ($\bar{g}^{(2)} \approx 1$). The value of gamma we use leads to a maximum ratio $g/\gamma \approx 200$, within reach of near future experiments in Circuit QED architectures \cite{houck2012chip}.

\section{Conclusions}
We have demonstrated the existence of a novel strongly correlated regime of `frozen' photons in optical resonator arrays with large Jaynes Cummings type nonlinearities. For a sufficiently large initial excitation of part of the resonator array, the photon dynamics are dramatically suppressed due to a very small overlap with propagating modes of the system. As little as 2 pumped photons per resonator are sufficient to observe signatures of `frozen' domains of photons, allowing access to the truly quantum few-excitation regime.


\begin{thebibliography}{10}
\newcommand{\enquote}[1]{``#1''}

\bibitem{angelakis2007photon}
D.~Angelakis, M.~Santos, and S.~Bose, \enquote{{Photon-blockade-induced Mott
  transitions and XY spin models in coupled cavity arrays},} Phys. Rev. A
  \textbf{76}, 31805 (2007).

\bibitem{hartmann2006strongly}
J.~Hartmann, \enquote{Strongly interacting polaritons in coupled arrays of
  cavities,} Nature Physics \textbf{2}, 849--855 (2006).

\bibitem{greentree2006quantum}
A.~Greentree, C.~Tahan, J.~Cole, and L.~Hollenberg, \enquote{{Quantum phase
  transitions of light},} Nature Physics \textbf{2}, 856--861 (2006).

\bibitem{tomadin2010signatures}
A.~Tomadin, V.~Giovannetti, R.~Fazio, D.~Gerace, I.~Carusotto, H.~T{\"u}reci,
  and A.~Imamoglu, \enquote{Signatures of the superfluid-insulator phase
  transition in laser-driven dissipative nonlinear cavity arrays,} Physical
  Review A \textbf{81}, 061801 (2010).

\bibitem{bamba2011origin}
M.~Bamba, A.~Imamo{\u{g}}lu, I.~Carusotto, and C.~Ciuti, \enquote{Origin of
  strong photon antibunching in weakly nonlinear photonic molecules,} Phys.
  Rev. A \textbf{83}, 021802 (2011).

\bibitem{leib2010bose}
M.~Leib and M.~Hartmann, \enquote{Bose--hubbard dynamics of polaritons in a
  chain of circuit quantum electrodynamics cavities,} New Journal of Physics
  \textbf{12}, 093031 (2010).

\bibitem{carusotto2009fermionized}
I.~Carusotto, D.~Gerace, H.~Tureci, S.~De~Liberato, C.~Ciuti, and
  A.~Imamo{\u{g}}lu, \enquote{{Fermionized photons in an array of driven
  dissipative nonlinear cavities},} Phys. Rev. Lett. \textbf{103}, 33601
  (2009).

\bibitem{gerace2009quantum}
D.~Gerace, H.~T{\"u}reci, A.~Imamo{\u{g}}lu, V.~Giovannetti, and R.~Fazio,
  \enquote{{The quantum-optical Josephson interferometer},} Nature Physics
  \textbf{5}, 281--284 (2009).

\bibitem{grujic2012repulsively}
T.~Grujic, S.~R. Clark, D.~Jaksch, and D.~G. Angelakis, \enquote{Repulsively
  induced photon super-bunching in driven resonator arrays,} arXiv preprint
  arXiv:1212.3012  (2012).

\bibitem{makin2009time}
M.~Makin, J.~Cole, C.~Hill, A.~Greentree, and L.~Hollenberg, \enquote{{Time
  evolution of the one-dimensional Jaynes-Cummings-Hubbard Hamiltonian},} Phys.
  Rev. A \textbf{80}, 43842 (2009).

\bibitem{schmidt2010nonequilibrium}
S.~Schmidt, D.~Gerace, A.~Houck, G.~Blatter, and H.~T{\"u}reci,
  \enquote{Nonequilibrium delocalization-localization transition of photons in
  circuit quantum electrodynamics,} Physical Review B \textbf{82}, 100507
  (2010).

\bibitem{longo2011few}
P.~Longo, P.~Schmitteckert, and K.~Busch, \enquote{Few-photon transport in
  low-dimensional systems,} Physical Review A \textbf{83}, 063828 (2011).

\bibitem{wong2011two}
M.~Wong and C.~Law, \enquote{Two-polariton bound states in the
  jaynes-cummings-hubbard model,} Physical Review A \textbf{83}, 055802 (2011).

\bibitem{hartmann2008migration}
M.~Hartmann and M.~Plenio, \enquote{{Migration of bosonic particles across a
  Mott insulator to a superfluid phase interface},} Phys. Rev. Lett.
  \textbf{100}, 70602 (2008).

\bibitem{mendoza2013dephasing}
J.~Mendoza-Arenas, T.~Grujic, D.~Jaksch, and S.~Clark, \enquote{Dephasing
  enhanced transport in non-equilibrium strongly-correlated quantum systems,}
  arXiv preprint arXiv:1302.5629  (2013).

\bibitem{perez2007matrix}
D.~Perez-Garcia, F.~Verstraete, M.~Wolf, and J.~Cirac, \enquote{Matrix product
  state representations,} Quantum Information \& Computation \textbf{7},
  401--430 (2007).

\bibitem{vidal2003efficient}
G.~Vidal, \enquote{{Efficient classical simulation of slightly entangled
  quantum computations},} Phys. Rev. Lett. \textbf{91}, 147902 (2003).

\bibitem{houck2012chip}
A.~Houck, H.~T{\"u}reci, and J.~Koch, \enquote{On-chip quantum simulation with
  superconducting circuits,} Nature Physics \textbf{8}, 292--299 (2012).
  
\end{thebibliography}
\end{document}